# Multiphase Monte Carlo and Molecular Dynamics Simulations of Water and CO$_2$ Intercalation in Montmorillonite and Beidellite

Meysam Makaremi,[1,2,3] Kenneth D. Jordan,[1,2] George D. Guthrie,[1,**] and Evgeniy M. Myshakin[1,4,*]

[1]*National Energy Technology Laboratory, 626 Cochrans Mill Road, Pittsburgh, Pennsylvania 15236*

[2]*Department of Chemistry, University of Pittsburgh, Pittsburgh, Pennsylvania 15260*

[3]*Department of Mechanical Engineering, University of Pittsburgh, Pittsburgh, Pennsylvania 15260*

[4]*AECOM, 626 Cochrans Mill Road, P.O. Box 10940, Pittsburgh, PA 15236*

### Abstract

Multiphase Gibbs ensemble Monte Carlo simulations were carried out to compute the free energy of swelling for Na-montmorillonite and Na-beidellite interacting with CO$_2$ and H$_2$O at pressure and temperature conditions relevant for geological storage aquifers. The calculated swelling free energy curves show stable monolayer and bilayer configurations of the interlayer species for Na-montmorillonite, while only the monolayer structure is found to be stable for Na-beidellite. The calculations show that CO$_2$ is intercalated into hydrated clay phases at concentrations greatly exceeding its solubility in bulk water. This suggests that expandable clay minerals are good candidates for storing carbon dioxide in interlayer regions. For Na-beidellite the CO$_2$ molecule distribution is mainly controlled by the position of the isomorphic substitutions, while for Na-montmorillonite the presence of the hydrated sodium ions plays an important role in establishing the CO$_2$ distribution.

[*]Corresponding author. E-mail address: Evgeniy.Myshakin@netl.doe.gov
[**]Present address: Los Alamos National Laboratory, P.O. Box 1663, Los Alamos, NM 87545

## 1. Introduction

Burning of fossil fuels releases large quantities of CO$_2$, a greenhouse gas, into the atmosphere, and there is compelling evidence that this is having a significant impact on the global temperature.[1] Sequestration of CO$_2$ in deep underground sedimentary layers is an approach widely suggested to reduce greenhouse gas emission.[2] In this approach, supercritical CO$_2$ (scCO$_2$) would be injected into deep rock formations. Successful sequestration depends on cap rocks being impermeable to CO$_2$. Cap rocks and host rock matrix can be enriched with swelling clays (smectites) such as montmorillonite and beidellite. The interaction of scCO$_2$ with hydrated smectites results in intercalation of $CO_2$ into interlayer regions.[3-6] Montmorillonite (MMT) and beidellite (BEI), which is a less common smectite clay mineral, are layered 2:1 dioctahedral phyllosilicates comprised of octahedral (*O*) sheets sandwiched between tetrahedral (*T*) sheets giving a TOT structure as show in Figure 1. In the absence of substitutions, the *T* sites are comprised of $SiO_4$ units and the *O* sites of $AlO_6$ units. Structural hydroxyl groups are also present in the coordination sphere of the octahedral metal. In 2:1 phyllosilicates one of three symmetrically unique octahedral sites is not occupied by a cation. MMT has predominant isomorphic substitutions of $Al^{3+}$ ions by $Mg^{2+}$ ions in the *O* sheet, while BEI samples tend to be dominated by isomorphic substitutions of $Si^{4+}$ ions by $Al^{3+}$ in the *T* sheets. Both types of substitution result in a net negative charge for the clay layer, which is balanced by interlayer cations such as $Na^+$, $Li^+$, and $Ca^{2+}$. Hydration of the cations causes swelling of smectites and an increase of the $d_{001}$-spacing.[7,8] In natural samples of MMT, some substitutions can also occur in the *T* and *O* sheets. Similarly, in natural samples of BEI, some substitutions occur in *O* sites. However, for simplicity we consider the ideal substitutions described above. This allows us to study the effect of location of the negative charges, exposed directly to interlayer species or screened by the silicon-oxygen polyhedra, on swelling behavior. Other factors affecting swelling behavior include the size and charge of the interlayer cations,[9,10] the positions of the hydroxyl groups in the octahedral layer with respect to the vacancies in octahedral sites, resulting in *cis*-vacant or *trans*-vacant smectites,[11] and the magnitude of negative charges on the clay sheets.[8] The expansion of smectites due to exposure to water and $CO_2$ improves the sealing properties of cap rocks, whereas shrinkage could lead to fracturing and undesirable release of injected $CO_2$ into the atmosphere.[4]

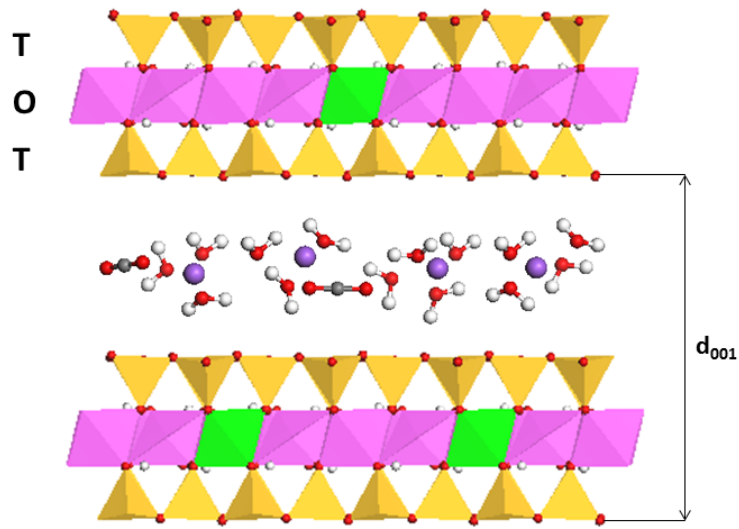

Figure 1. Smectite structure containing TOT sheets, counterions, and intercalated water and $CO_2$ molecules. The color coding of atoms is: purple, red, white, grey, yellow, pink and green for Na, O, H, C, Si, Al and Mg, respectively.

The swelling of smectite clays due to water intercalation is well understood from previous studies with emphasis on montmorillonite as one of the most abundant smectite minerals,[12-20] while beidellite has received less attention.[21-23] Previous calculations on MMT indicate that at equilibrium water can form a monolayer with basal $d_{001}$-spacings ranging from 11.5 to 12.5 Å (the 1W hydration state) or a bilayer with basal $d_{001}$-spacings ranging from 14.5 to 15.5 Å (the 2W hydration state).[20,21,24,25] The relative stability of the 1W and 2W hydration states depends on the type of interlayer ions.[26] In this work "1W" and "2W" are used, respectively, to refer to the monolayer and bilayer configurations of interlayer molecules for both the pure water and for $CO_2/H_2O$ mixtures.

Smith used Monte Carlo simulations to investigate the swelling of hydrated Cs-MMT and to calculate the immersion energy to establish the hydration state corresponding to the global free energy minimum.[26] The minimum was found to be consistent with the experimental result.[27,28] In subsequent work, Young and Smith[10] calculated the structural parameters for three different MMTs and found that for Cs-MMT-$H_2O$ formation of a single water layer in the interlayer region leads to the most stable structure, while for Na-MMT-$H_2O$ and Sr-MMT-$H_2O$ the 2W hydration state was predicted to be more stable than 1W. The energetic contributions to the swelling free energy were calculated by Whitley and Smith,[29] who reported that hydration of the counter-ions plays a dominant role in the swelling behavior of MMTs.

Interaction of carbon dioxide with hydrated smectites is a complex phenomenon.[6,30] For example, exposure of dry sc$CO_2$ to MMT samples in the 2W and higher hydration states can induce a collapse of the $d_{001}$-spacing to that of the 1W state.[6,30] On the other hand, interaction of MMT samples with wet sc$CO_2$ promotes swelling of clays to either 1W or 2W hydration states.[3,4,6,30]

The interaction of carbon dioxide and simple hydrocarbons with hydrated swelling clays has been studied using computer simulation methods.[21-23,31-42] Molecular dynamics (MD) simulations have been engaged to characterize the swelling process as well as the structure and transport of interlayer species.[36,39-42] Botan et al.[43] applied grand canonical Monte Carlo (μVT-MC) simulations to study the effect of water and $CO_2$ intercalation on Na-MMT swelling at prevailing (P, T) conditions of an underground formation. They considered basal $d_{001}$-spacings over the range of 12 to 17 Å that excluded important $d_{001}$-spacings corresponding to sub-1W hydration levels. In addition, we note that the µVT-MC method needs precise calculation of chemical potentials for the binary $CO_2/H_2O$ mixture which is quite challenging.

The goal of the present study is to characterize intercalation of pure carbon dioxide, pure water, and $CO_2/H_2O$ mixtures into Na-MMT and Na-BEI at two (*P*, *T*) conditions, one relevant to the geological carbon storage *via* injection of $scCO_2$ and the other corresponding to exposure of clay to $CO_2$ in the gas phase. Na-MMT and Na-BEI have different localization of the negative charges in their mineral layers allowing us to study whether this is an important factor in $CO_2$ uptake. The simulations employed the Gibbs ensemble Monte Carlo (GEMC) method[44,45] in which knowledge of chemical potentials is not required to equilibrate non-interacting systems representing different phases of interest. In addition, molecular dynamics simulations in the canonical ensemble were used to analyze the properties of equilibrium water-$CO_2$ mixtures in the interlayer region of clays.

## 2. Computational Details

### 2.1 Swelling thermodynamics

The thermodynamics of swelling of layered materials was outlined by Diestler *et al.*[46] and Bordarier *et al.*[19] As shown in those publications, the internal energy of a slit micropore can be expressed as:

$$dU = TdS + \mu dN + \sigma_{xx}s_y s_z ds_x + \sigma_{yy}s_z s_x ds_y + \sigma_{zz}s_x s_y ds_z \quad (1)$$
$$+ \tau_{zx}s_x s_y dl_x + \tau_{zy}s_x s_y dl_y.$$

where *T*, *S*, *µ*, *N*, $\sigma_{ii}$, $\tau_{ij}$ (*i,j* = *x,y,z*) are, respectively, the temperature, entropy, chemical potential, number of molecules, and normal and shear stress terms. $s_i$ and $l_i$ are Cartesian components of the micropore depicted in Figure 2.

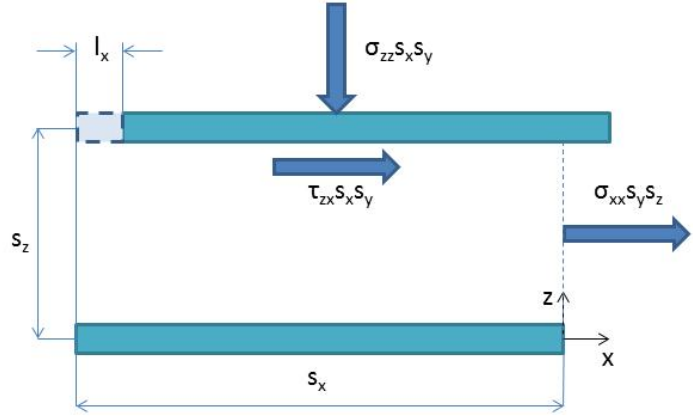

Figure 2. Slit micropore structure including position and force components.

Taking into account only the normal stress term in the z direction, the free energy, F, and isostress free energy, Φ, are defined as

$$F = U - TS - \mu N, \tag{2}$$

$$\Phi = F - \sigma_{zz} A s_z, \tag{3}$$

where A is the surface area of the slit. Thus the difference in the swelling free energy for change in wall separation $\Delta s_z$ can be written as

$$\Delta \Phi = \Delta F - \sigma_{zz} A \, \Delta s_z \tag{4}$$

By using equations 1 and 4, eliminating the *x* and *y* directions, and also considering the fact that *P* and *T* are fixed at phase equilibrium, gives

$$\Delta \Phi = A \int_{s_z^0}^{s_z} (\sigma_{zz}(s'_z) - \sigma'_{zz}) ds'_z, \tag{5}$$

where $\sigma'_{zz}$ is the applied constant stress. By eliminating the isotropic part of the stress (bulk pressure) from the stress terms, the equation can be converted into a pressure based equation

$$\Delta \Phi = -A \int_{s_z^0}^{s_z} (P_{zz}(s'_z) - P'_{zz}) ds'_z. \tag{6}$$

where $P_{zz}(s'_z)$ and $P'_{zz}$ are the slit normal (disjoining) and the applied pressure, respectively. Equation 6 has been used by Smith[26] and by Botan *et al.*[43] to calculate the free energy of swelling as a function of the distance between layers.

## 2.2. Gibbs Monte Carlo Method

The use of the multiphase Gibbs approach to study equilibrium of multi-component systems is well-documented in literature.[47-52] Lopes and Tildesley[47] evaluated the multicomponent GEMC approach to calculate two-phase, three-phase and four-phase equilibria. Kristof *et al.*[50] applied the approach to the three-component, $CO_2$-water-methanol system at high pressure conditions, obtaining good agreement between the results of their simulations and the experimental data.

The Gibbs ensemble Monte Carlo (GEMC) method[44,45] uses separate boxes for each phase to calculate the phase equilibrium for a target ($P,T$) condition and to achieve equivalence of chemical potentials of interacting species. The algorithm employs three basic trial moves: particle displacement, particle exchange, and volume change. The displacement moves include a transitional or moves in a single box to satisfy temperature equilibrium. The exchange moves involve deletion of a particle in one box and its insertion into another box, to assure chemical potential equilibration. The volume moves expand or contract boxes to fulfill the pressure equilibration. Figure 3 shows the simulation setup with three boxes representing water, carbon dioxide, and clay phases.

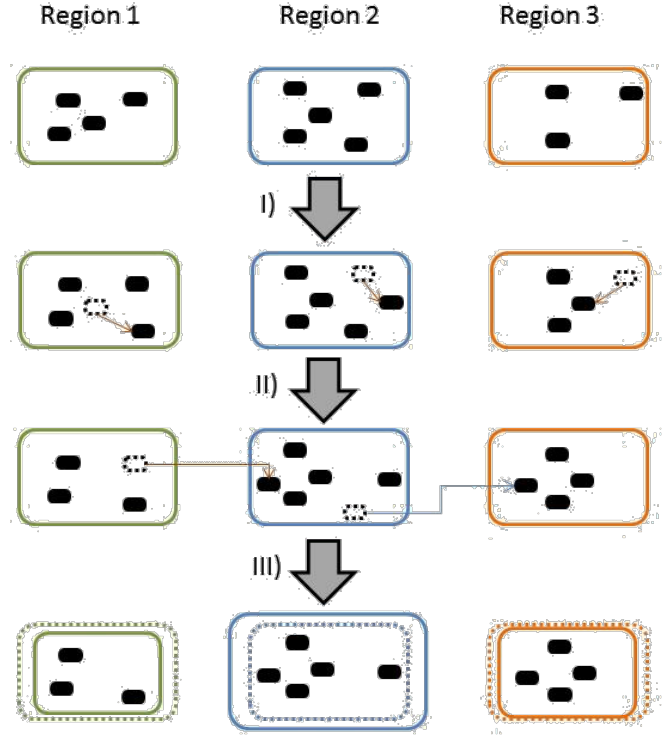

Figure 3. Illustration of Gibbs Monte Carlo method including three phases and three types of trial moves: i) displacement move, ii) exchange move, and iii) volume move.

**2.3. Models and Force Fields**

In this work, we used Na-MMT and Na-BEI clay structures with the stoichiometric chemical formulas $Na_{0.75}Mg_{0.75}Al_{3.25}(OH)_4(Si_4O_{10})_2$ and $Na_{0.75}Al_4(OH)_4(Al_{0.375}Si_{3.625}O_{10})_2$, respectively, where each clay TOT layer bears a negative charge -0.75 $|e|$ per unit cell that it is counterbalanced by interlayer $Na^+$ cations. The isomorphic substitutions in the TOT sheets were placed in such a way that they show a repeated pattern in the directions of *a* and *b* lattice vectors. The Al-O-Al linkage in the *T* sheets was avoided according to Loewenstain's rule.[53] The clay structure is described by the Clayff force field,[25] while the rigid SPC[54] and the rigid EPM2-based model of Cygan et al.[35] are employed for water and $CO_2$, respectively. The total potential energy can be written as:

$$E_{Non-bond} = E_{VDW} + E_{Coul} + E_{Stretch}, \qquad (7)$$

with

$$E_{VDW} = 4\epsilon_{ij}[(\frac{\sigma_{ij}}{r_{ij}})^{12} - (\frac{\sigma_{ij}}{r_{ij}})^6], \quad E_{Coul} = \frac{q_i q_j e^2}{4\pi\epsilon_0 r_{ij}},$$

where atoms *i* and *j* are separated by a distance $r_{ij}$, and $q_i$ is the charge on the atom *i*, e is the elementary charge of an electron, $\epsilon_0$ is the vacuum permittivity, and $\epsilon_{ij}$ and $\sigma_{ij}$ are the Lennard-Jones energy and distance parameters, respectively. The parameters for unlike atoms were calculated by using the Lorentz-Berthelot classical combining rules:[55]

$$\sigma_{ij} = \tfrac{1}{2}(\sigma_i + \sigma_j), \quad \epsilon_{ij} = \sqrt{\epsilon_i \epsilon_j}. \tag{8}$$

The bond stretch ($E_{\text{Stretch}}$) potential term is employed only for the hydroxyl groups of the clay layer and is taken to be harmonic.

## 2.4. Simulation methodology

The NPT-GEMC method was applied to the ternary clay-$H_2O$-$CO_2$ and binary $H_2O$-$CO_2$ systems at two ($P$, $T$) conditions, (25 bar, 348.15 K and 125 bar, 348.15 K). The binary system was included to allow comparison of the results of our calculations with previous experimental[56,57] and computational[43,58] data, and also to provide a reference for the subsequent simulations of the ternary system.

In the MC simulations of the Na-MMT and Na-BEI phases, 4 x 3 x 2 (20.8 x 27.6 x *2h* Å) supercells, requiring 18 counterions for neutrality, were employed. Calculations were carried out for $d_{001}$-spacings, $h$, ranging from 9.5 to 18 Å in steps of 0.5 Å to cover separations ranging from nearly dry clays to the interlayer distances corresponding to the 3W hydration state. For spacings >3W, the behavior of interlayer species becomes similar to that in the bulk.[12-20] The MC simulations were carried out for $2 \times 10^5$ equilibration cycles, followed by $1.5 \times 10^5$ production cycles. Each cycle consisted of 818 attempted moves, including volume exchange, configurational-bias interbox and intrabox molecule transfer, molecule regrowth, translational, and rotational moves.[59-61] Exploratory calculations were carried out to determine optimal probabilities of the various types of moves with the initial values being 0.5% volume exchange, 39.5% configurational-bias consisting of both inter- and intra-box moves, 10% regrowth, and 50% displacement, including translational and rotational moves. During the exploratory simulations, the probabilities of different moves and the components in the boxes were adjusted to ensure that the calculated chemical potentials of water and $CO_2$ in the various phases agreed to within the standard deviations. The positions of the atoms in the clay layers were held fixed, while the $Na^+$ ions were allowed to move within the interlayer space. The $H_2O$ and $CO_2$ molecules were allowed to migrate between the "$H_2O$-rich", "$CO_2$-rich," and clay boxes. The cut-off radii for the non-bonded van der Waals interactions and for the real-space part of the Ewald summation of the electrostatics were chosen to be 9.5 Å, with switching distances starting from 6.5 Å in the Ewald summation. Standard deviations of all ensemble averages were determined by the standard block average technique.[62] The NPT-GEMC simulations were carried out using the Towhee simulation package.[63]

In addition to the GEMC simulations described above, molecular dynamics simulations in the canonical ensemble (NVT-MD) were carried out to analyze various dynamical properties of the molecules in the clay phase. These simulations used a triclinic 8 x 4 x 4 supercell with 5120 atoms constituting the mineral portion of the clay phase. The negative charge introduced by the isomorphic substitutions is compensated by 96 sodium ions residing in all interlayers. The carbon dioxide and water content (mole fractions) were taken to correspond to those obtained for the minimum free-energy structures from the Monte Carlo simulations. Equilibration runs of 1 $ns$ were carried out, followed by 5 $ns$ production runs with temperature controlled by a Nóse-Hoover thermostat[64,65] with a relaxation time of 2 $ps$. The cut-off radii for the non-bonded van der Waals interactions and for the Ewald summation of the electrostatics were chosen to be 9.5 Å, with the switching distance for the latter starting from 8.5 Å. Due to the use of cut-offs for the LJ interactions, long-range dispersion corrections for energy and pressure were applied. The leap-frog algorithm[66] was used to update positions every 1 $fs$. The MD simulations were carried out using the GROMACS package.[67]

The radial distribution function (RDF) for species $B$ around species $A$ was calculated using:

$$G_{A-B}(r) = \frac{1}{4\pi\rho_B r^2} \frac{\Delta N_{A-B}}{\Delta r}, \tag{9}$$

where $\rho_B$ is the number density of species $B$, and $\Delta N_{A-B}$ is the average number of type $B$ particles lying in the region $r$ to $r + \Delta r$ from a type $A$ particle. $N_{A-B}$ is the coordination number for $B$ around $A$.

The diffusion coefficients of the interlayer species were calculated using the Einstein relation and equilibrated atomic trajectories from the NVT ensemble simulations with 1 $ps$ sampling to evaluate the mean square displacement of the species of interest:

$$D = \frac{1}{6N_m t} \left\langle \sum_{j=1}^{N_m} [r_j(t) - r_j(0)]^2 \right\rangle, \tag{10}$$

where $N_m$ is the number of the species averaged over, and $r_j(t)$ is the center-of-mass position of the $j$th species at time $t$. The averages were over 5 $ns$ trajectories. The diffusion coefficients were derived from the linear slope of the mean square displacement as a function of the simulation time. Different restart points in the analysis were used to monitor convergence.

## 3. Results and Discussion

### 3.1. Monte Carlo simulations

Table 1 compares the $CO_2$ mole fractions and the densities of the $CO_2$-rich and water-rich phases of the two-phase water/$CO_2$ and three-phase water/$CO_2$/clay systems. In the latter case, calculated results are reported for both Na-MMT and Na-BEI at $d_{001}$ = 12.5 Å.

Table 1: Mole fractions of $CO_2$ ($X_{CO_2}$) and densities ($\rho$) of $H_2O$ and $CO_2$ in multiphase systems from previous studies and current work ($T$=348.15 K).

| system | P [bar] | $\rho$ [g/cm$^3$] | | $X_{CO_2}$ | | |
| --- | --- | --- | --- | --- | --- | --- |
| | | $H_2O$ | $CO_2$ | $H_2O$ | $CO_2$ | clay ($d_{001}$=12.5 Å) |
| $H_2O$-$CO_2$ (Experiment)[57] | 25.3 | | | 0.0054 | 0.9818 | |
| | 126.7 | | | | 0.9915 | |
| $H_2O$-$CO_2$ ($\mu$VT-MC)[43] | 24.2 | 0.934(3) | 0.039(1) | 0.004(1) | 0.982(1) | |
| | 122.4 | 0.940(5) | 0.298(3) | 0.014(1) | 0.994(1) | |
| $H_2O$-$CO_2$ (GEMC) | 25 | 0.939(4) | 0.040(1) | 0.004(1) | 0.980(4) | |
| | 125 | 0.947(4) | 0.314(7) | 0.016(3) | 0.994(2) | |
| MMT-$H_2O$-$CO_2$ (GEMC) | 25 | 0.937(5) | 0.040(1) | 0.004(1) | 0.980(2) | 0.014(2) |
| | 125 | 0.947(5) | 0.317(5) | 0.016(3) | 0.994(1) | 0.056(2) |
| BEI-$H_2O$-$CO_2$ (GEMC) | 25 | 0.933(4) | 0.040(1) | 0.005(2) | 0.983(2) | 0.022(2) |
| | 125 | 0.945(4) | 0.321(5) | 0.017(3) | 0.993(1) | 0.036(3) |

For the $CO_2$/water system, the calculated mole fractions are in excellent agreement with the results of experimental measurements and with the results of grand canonical Monte Carlo simulations by Botan *et al.*[43] Similar mole fractions of $CO_2$ are found in the $CO_2$ and water phases of the three-phase systems as for the water/$CO_2$ two-phase system. Also, the densities calculated for the water-rich and $CO_2$-rich phases are nearly identical for the two-phase and three-phase systems. Most significantly, the concentration of carbon dioxide in the clay phase coexisting at equilibrium with the $CO_2$/$H_2O$ binary system greatly exceeds that in the bulk water-rich phase.

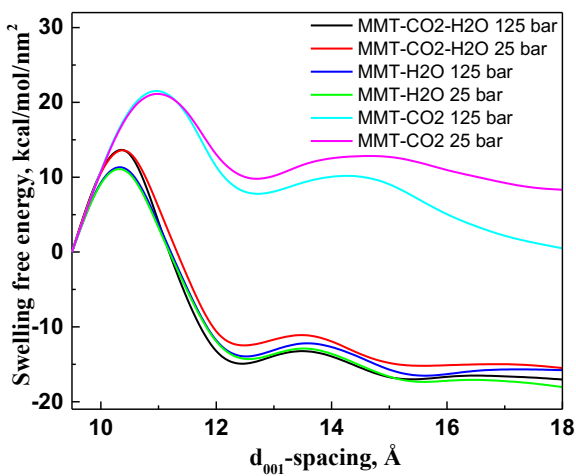 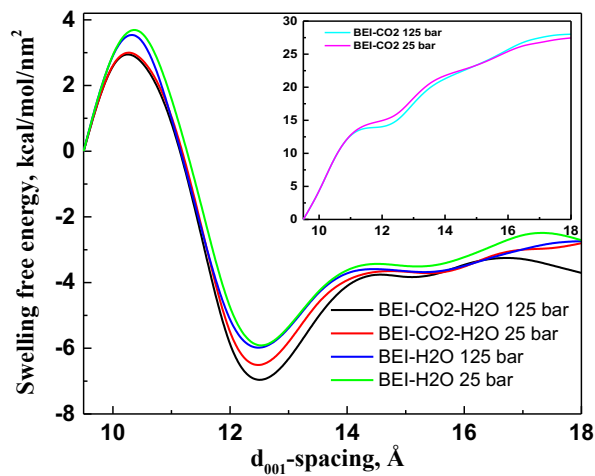

Figure 4. Calculated swelling free energy of Na-MMT as a function of the basal $d_{001}$-spacing for intercalation of pure $CO_2$, pure $H_2O$, and the $H_2O$-$CO_2$ mixture at $P$ = 25 and 125 bar, $T$ = 348.15 K

Figure 5. Calculated swelling free energy of Na-BEI as a function of the basal $d_{001}$-spacing for intercalation of pure $CO_2$ (insert), pure $H_2O$, and the $H_2O$-$CO_2$ mixture at $P$ = 25 and 125 bar, $T$ = 348.15 K.

The calculated swelling free energies as a function of the $d_{001}$-spacing for pure $CO_2$, pure $H_2O$ and mixed $H_2O$-$CO_2$ intercalation into Na-MMT are depicted in Figure 4. Similar profiles for Na-BEI-$H_2O$ and Na-BEI-$CO_2$-$H_2O$ are given in Figure 5. The smallest (9.5 Å) $d_{001}$-spacing in the figures corresponds to dry clays in their 0W hydration state, characterized by the absence of intercalated molecules and provides a common origin for the swelling free energy curves. From the profiles, it is seen that in the absence of water, intercalation of $CO_2$ is highly unfavorable thermodynamically. For Na-MMT-$CO_2$ the minimum at 12.5 Å is about 10 kcal/mol/nm² less stable than that at 9.5 Å, *i.e.* the 0W hydration state. This result is consistent with experimental studies that show that carbon dioxide does not intercalate into MMTs in the absence of water.[3,30,68] Our calculations also show that, in the absence of water, intercalation of $CO_2$ into Na-BEI is even more unfavorable than into Na-MMT.

For the intercalation of pure water into montmorillonite, two minima corresponding to 1W and 2W are identified with energies below that of 0W. The stability of the 1W and 2W states is a consequence of ion hydration and hydrogen bonding between the intercalated water molecules and the basal oxygen atoms.[8] The existence of two stable hydration states of Na-MMT is supported by previous studies.[27,69-72] The present calculations predict the 2W hydration state to be more stable than the 1W hydration state as found in earlier MC simulations[29,73] For pure water in Na-MMT, the barrier for going from the 0W to the 1W state is calculated to be about 10 kcal/mol/nm$^2$ and is relatively insensitive to pressure. In contrast to Na-MMT-H$_2$O, Na-BEI-H$_2$O does not have a stable 2W state (Figure 5). The inability of Na-BEI-H$_2$O to swell beyond 1W has been documented in the literature,[74] and factors responsible for this behavior will be discussed later in Section 3.2.

The swelling free energy curves for the clay systems with intercalated binary H$_2$O-CO$_2$ mixtures display trends similar to those for pure water (Figures 4 and 5). For both Na-MMT-CO$_2$-H$_2$O and Na-BEI-CO$_2$-H$_2$O, the free energy minima at $d_{001}$-spacings corresponding to the 1W hydration states are deeper and the barriers separating 0W and 1W are several kcal/mol/nm$^2$ lower than those for swelling induced by pure water. For Na-MMT-CO$_2$-H$_2$O the free energy barriers between the minima at 12.5 and 15 Å spacing is calculated to be only about 3 kcal/mol/nm$^2$ with the minimum at 15.5 Å being more stable. This indicates that in geological aquifers, both mono- and bilayer arrangements of CO$_2$-H$_2$O may occur in clays rich with MMT. Increasing the pressure is found to stabilize the free-energy minima near $d_{001}$ = 12.5 Å (1W) of both clay systems.

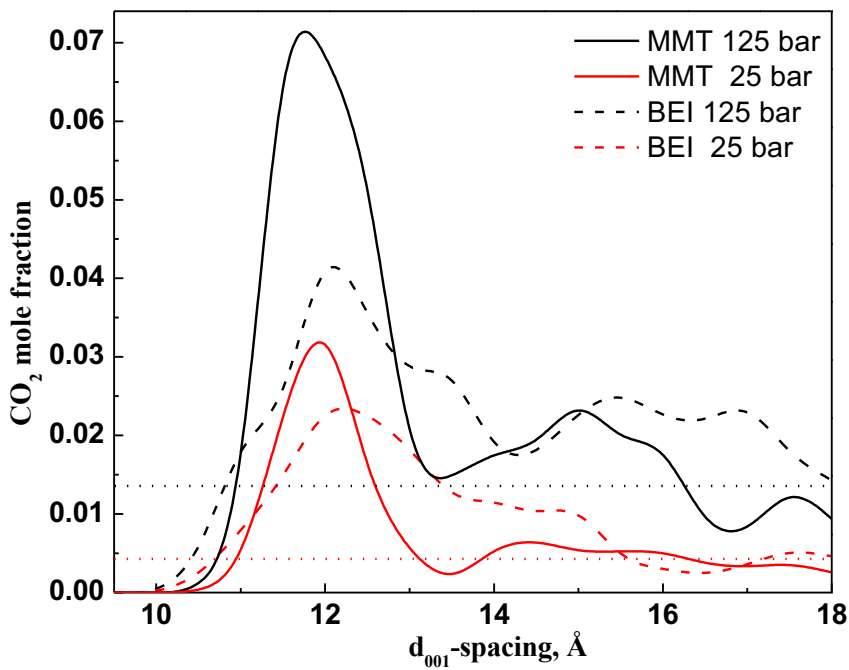

Figure 6. Equilibrium CO$_2$ mole fraction as a function of basal $d_{001}$-spacing for Na-MMT-CO$_2$-H$_2$O and Na-BEI-CO$_2$-H$_2$O at $T$= 348.15 K and $P$ = 25 or 125 bar. The horizontal dotted lines represent the CO$_2$ mole fractions in the bulk H$_2$O-rich phase at $P$ = 125 bar (black) and $P$ = 25 bar (red), both at T=348.15 K.

The calculated mole fraction of $CO_2$ as a function of the basal $d_{001}$-spacing of Na-MMT and Na-BEI is reported in Figure 6. In both systems, there is a pronounced peak in the $CO_2$ mole fraction at a $d_{001}$-spacing corresponding to the 1W state and a second weaker peak (or shoulder) at a distance corresponding to 2W. Experimental studies of Na- and Ca-MMT[3,68,75] have shown that the maximum expansion of the interlayer due to $CO_2$ exposure occurs at the sub 1W hydration state. Increasing the pressure causes an elevation of the $CO_2$ concentration in the interlayer region, with the increase being greater for Na-MMT than for Na-BEI. For both Na-MMT-$CO_2$-$H_2O$ and Na-BEI-$CO_2$-$H_2O$ the concentration of $CO_2$ in the interlayer region greatly exceeds that in the bulk water-rich phase. For Na-MMT-$CO_2$-$H_2O$, the calculated mole fractions of $CO_2$ in the interlayer region obtained in the present study in fairly good agreement with those reported by Botan *et al.*,[43] although the $CO_2$ concentrations predicted in this work are lower. The different concentrations may be the result of the different force fields used in the two theoretical studies.

## 3.2. Molecular dynamics simulations

The Na-MMT-$CO_2$-$H_2O$ and Na-BEI-$CO_2$-$H_2O$ systems at $d_{001}$-spacing of 12.5 Å (1W) and 15.5 Å (2W) equilibrated in the GEMC simulations were further characterized by molecular dynamics using the NVT ensemble. In particular, the MD simulations were used to calculate the structural and transport properties of the interlayer species. The Na-BEI-$CO_2$-$H_2O$ system at $d_{001}$ = 15.5 Å does not correspond to a free energy minimum and is considered primarily for comparison with Na-MMT-$CO_2$-$H_2O$ at the same $d_{001}$-spacing.

### 3.2.1. Structural details of the interlayer species

Figure 7 reports for Na-MMT and Na-BEI the calculated atomic density profiles for water oxygen ($O_w$), carbon ($C_{CO2}$), and sodium ions ($Na^+$) at $d_{001}$-spacings equal to 12.5 and 15.5 Å. The profiles were computed along the *z* axis (perpendicular to the clay surface) and averaged over the interlayer region of the simulation box using a 3 *ns* production time frame. For each species of interest, the density profiles computed for P = 25 and 125 bars are close agreement, and only the *P* = 125 bar data are shown in the figure. The $O_w$ and $Na^+$ density profiles for pure water are qualitatively similar to those for the $H_2O$/$CO_2$ mixtures and, therefore, are not reported.

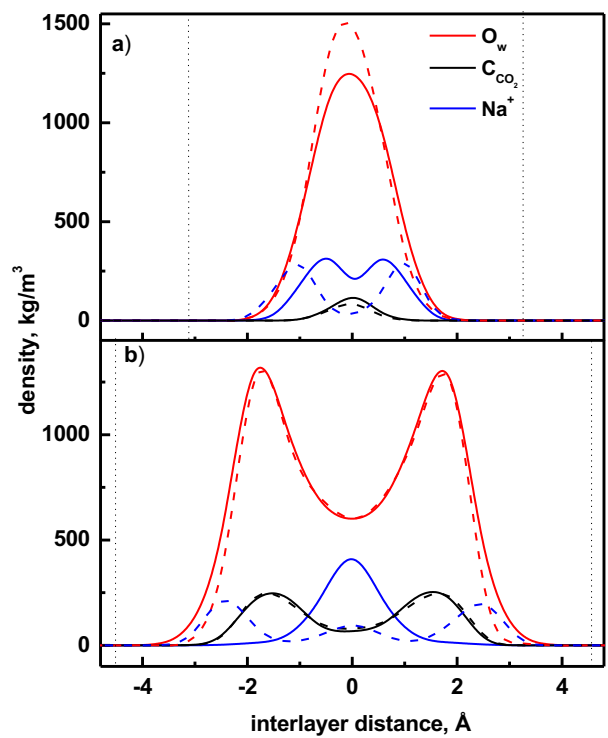

Figure 7. Density profiles of $O_w$, $C_{CO_2}$, and $Na^+$ along the interlayer distance perpendicular to the internal surfaces of Na-MMT-$CO_2$-$H_2O$ (solid lines) and Na-BEI-$CO_2$-$H_2O$ (dashed lines). The distributions of the intercalated species are reported for $d_{001}$-spacings of **a)** 12.5 Å and **b)** 15.5 Å For **b)** the densities of $C_{CO_2}$ are magnified by a factor of 10 for clarity. Vertical black dotted lines designate the planes of the basal oxygen atoms.

For both Na-MMT and Na-BEI, the density profiles for $O_w$ and $C_{CO_2}$ display a single peak in the case of $d_{001}$ = 12.5 Å and two peaks in the case of $d_{001}$ = 15.5 Å. For Na-MMT-$CO_2$-$H_2O$ the compositions of carbon dioxide and water in the monolayer and the bilayer structures are ~4.1 $H_2O$ / 0.3 $CO_2$ and ~8.9 $H_2O$ / 0.2 $CO_2$ per unit cell, respectively. These numbers are in reasonable agreement with the results of grand canonical MC simulations of Botan *et al.*[43] (monolayer: 3.7 $H_2O$ / 0.6 $CO_2$ and bilayer: 8.6 $H_2O$ / 0.4 $CO_2$ per unit cell). For Na-BEI-$CO_2$-$H_2O$ our simulations give ~4.4 $H_2O$ / 0.2 $CO_2$ and ~8.6 $H_2O$ / 0.2 $CO_2$ per unit cell for the monolayer and bilayer arrangement, respectively.

In an earlier paper,[36] we reported the $d_{001}$-spacing of Na-MMT as a function of the number of water molecules at fixed numbers of $CO_2$ molecules per unit cell in order to estimate the effect of $CO_2$ intercalation on the interlayer distance of Na-MMT-$CO_2$-$H_2O$. It was deduced that the addition of 0.2-0.3 $CO_2$ molecules per unit cell led to an expansion of the $d_{001}$-spacing by ~0.5-0.6 Å. In experimental studies performed at elevated ($P,T$) conditions the maximum expansion of Na-MMT upon interaction with dry $scCO_2$ was found at an initial $d_{001}$-spacing below 11.5 Å, which corresponds to a sub-1W state of hydrated Na-montmorillonite and was capped at a $d_{001}$-spacing equal to about 12.5 Å suggesting that expansion due to interaction with $scCO_2$ can be as much as ~1 Å.[3,6] Further, Loring et al.[75] estimated that the maximum $CO_2$ absorption in the interlayer of Na-exchanged MMT occurs at water concentrations close to the transition between 0W and 1W hydration states. Thus, the values of the equilibrium concentrations of intercalated $CO_2$ predicted in this work appear to be somewhat underestimated.

In contrast to the density profiles of $O_w$ and $C_{CO2}$, the density profiles of $Na^+$ are quite different for Na-MMT-$CO_2$-$H_2O$ and Na-BEI-$CO_2$-$H_2O$. Since the major difference in these two phyllosilicates is the location of the negative charge in the mineral layers, we conclude that this is an important factor determining the sodium ion distributions. For the $d_{001}$ = 12.5 Å case, the $Na^+$ density distribution is bimodal for both Na-MMT and Na-BEI. However, for Na-BEI, the positions of the peaks are shifted closer to the basal surfaces as a result of preferential coordination of sodium cations at the middle of the hexagonal rings formed by the silicon-oxygen polyhedra with isomorphic substitutions.[36] Effectively, the peak shift reflects a change from predominantly $Na^+$ "outer-sphere" coordination with water in case of Na-MMT to "inner-sphere" adsorption at the internal clay surfaces for Na-BEI where basal oxygens participate in the first solvation shell of sodium ions. Intercalation of carbon dioxide disrupts the water network[32] and also causes the $Na^+$ ions to be displaced toward the clay surfaces.[36]

In the 2W state of the Na-MMT-$CO_2$-$H_2O$ system, the $Na^+$ distribution peaks at the middle of interlayer. However, for Na-BEI-$CO_2$-$H_2O$, it is split into three peaks, with two more pronounced peaks corresponding to sodium ions adsorbed at the surfaces and a weaker central peak corresponding to $Na^+$ ions more fully solvated by water molecules. The surface-adsorbed $Na^+$ ions have in their first "solvation" shell both basal oxygens of the ditrigonal rings and interlayer water molecules. Additional information on the solvation structure of the ions can be gained by examining the radial distribution functions (RDFs) for atomic pairs of interest. Figure 8 reports the RDFs for the sodium ion – water oxygen pair computed for Na-MMT-$CO_2$-$H_2O$ and Na-BEI-$CO_2$-$H_2O$ at monolayer and bilayer arrangement and P = 125 bar. Figure 9 depicts the corresponding RDFs for the sodium ion – basal oxygen pair. Additionally, cumulative number RDFs, *i.e.* the average number of atoms within a distance, *r*, were calculated.

For the monolayer arrangements of Na-MMT-$CO_2$-$H_2O$ of Na-BEI-$CO_2$-$H_2O$, the $Na^+$ ions are hydrated on average by 4.0 and 3.6 water molecules, respectively. On the other hand, the average number of basal O atoms directly bonded to the $Na^+$ ions is 1.2 for Na-MMT, but 1.9 for Na-BEI. Thus, the total $Na^+$ coordination numbers in the monolayer case is about 5.3 in both cases. For the bilayer ($d_{001}$=15.5 Å) of Na-MMT-$CO_2$-$H_2O$ the Na+ ions tend to be coordinated by water molecules alone, achieving on average a coordination number of 5.5. On the other hand, for Na-BEI-$CO_2$-$H_2O$ in the 2W state, the average $Na^+$ coordination number, 5.7, derives from 4.3 of water oxygens and 1.4 of basal oxygens (Figure 9). A simulation of sodium ions in bulk water using the same (*P,T*) conditions and force fields as used in the clay simulations gives an average $Na^+$ coordination number 5.6 for the first hydration shell, essentially the same as that calculated for $Na^+$ in the bilayer in Na-MMT. Experimental studies of $Na^+$ in bulk water lead to $Na^+$ coordination numbers in the range 4.4-5.4[76,77] consistent with the results of CPMD simulations[78] which give an average coordination number of 4.9 at 300K. Thus, it appears that our simulations tend to slightly overestimate the average number of water molecules coordinating the $Na^+$ ions. This probably reflects a deficiency of the force field used in the simulations.

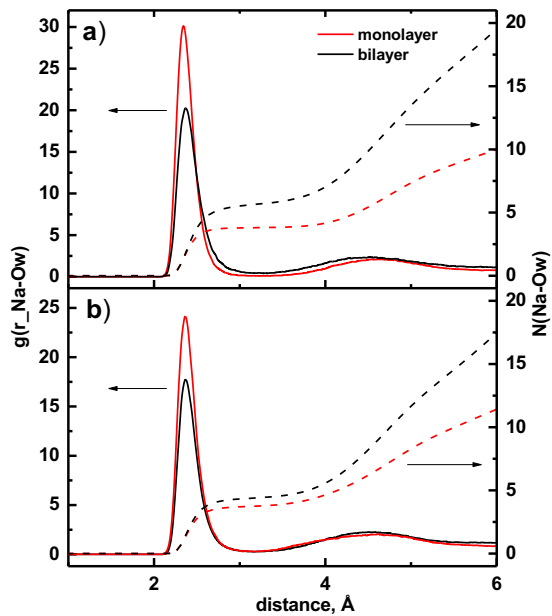
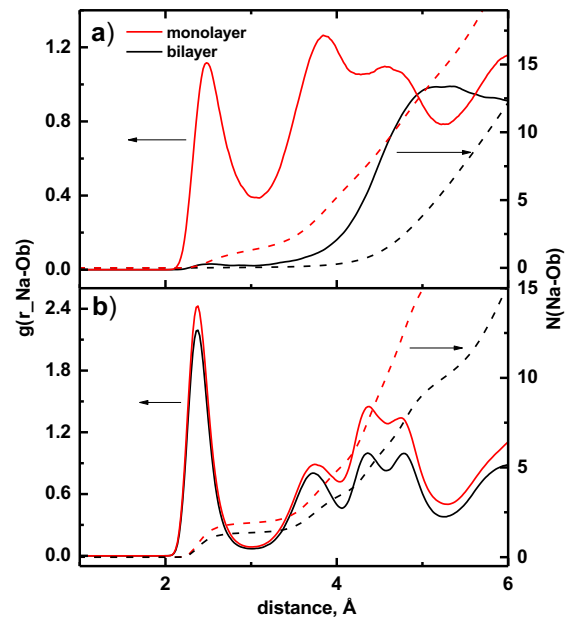

Figure 8. Radial distribution functions (solid curves) and cumulative number RDFs (dashed curves) of the $Na^+$-$O_w$ pair for **a)** Na-MMT-$CO_2$-$H_2O$ and **b)** Na-BEI-$CO_2$-$H_2O$ at P = 125 bar and T=348.15 K. Dotted lines designate the cumulative number RDF, *i.e.* the average number of particles within a $Na^+$-$O_w$ distance.

Figure 9. Radial distribution functions (solid curves) and cumulative number RDFs (dashed curves) of the $Na^+$-$O_b$ pair for **a)** Na-MMT-$CO_2$-$H_2O$ **b)** Na-BEI-$CO_2$-$H_2O$ at P = 125 bar and T=348.15 K. the cumulative number RDF, *i.e.* the average number of particles within a $Na^+$-$O_b$ distance.

In Na-BEI the strong electrostatic interaction between ions and the surfaces makes it more favorable for the cations to be adsorbed at a surface than to be located at the middle of the interlayer. In Na-MMT, in which the negative charges are screened by the *T* sheets, the Na+ ions are more stable near the center of the bilayer where they are essentially fully hydrated. Suter *et al.*[21] explored the thermodynamics of $Li^+$, $Na^+$ and $K^+$-montmorillonote and beidellite using *ab initio* molecular dynamics. These authors found that smectites with the free energy minimum in the middle of the interlayer for bilayer swell to the 2W hydration state with the ions forming a full hydration sphere. In Na-beidellite the position of the lowest energy region remains unchanged from monolayer to bilayer indicating that swelling would stop at monolayer. At that position, basal oxygens form a part of sodium hydration sphere. Thus, the interplay between the hydration of counter-balancing ions and the attraction to the basal surface determines the swelling behavior of smectites. Localization of charge in the *O* clay sheets supports swelling to 2W (and higher) while localization of the negative charge in the *T* sheets favors 1W (Figure 4 and 5).

We now consider the effects of layer charge localization and the interlayer ions on the $CO_2$-$H_2O$ distributions. Figure 10 shows the calculated RDFs of carbon-water oxygen (C-$O_w$) and $CO_2$ oxygen – water hydrogen ($O_{CO2}$-$H_w$) pairs for Na-MMT-$CO_2$-$H_2O$ for both monolayer and bilayer conditions at $P = 125$ bar. The corresponding RDFs for Na-BEI-$CO_2$-$H_2O$ (monolayer and bilayer) display qualitatively similar shapes and are not shown here. The C-$O_W$ RDF displays only a weak shoulder near 3.1 Å, corresponding to coordination of water to carbon of $CO_2$. The shoulder is more pronounced in the bilayer than in the monolayer case, consistent with a larger number of water oxygens surrounding $CO_2$ molecules in the former. However, the RDF of carbon – basal oxygens indicates that the coordination number of basal oxygens is larger for the monolayer than for the bilayer, so on average the $CO_2$ molecules are surrounded by similar numbers of oxygens atoms in the monolayer and bilayer. This result is consistent with a prior analysis of the nearest-neighbor coordination numbers for $CO_2$ carbon atoms in hydrated Na-MMT.[43]

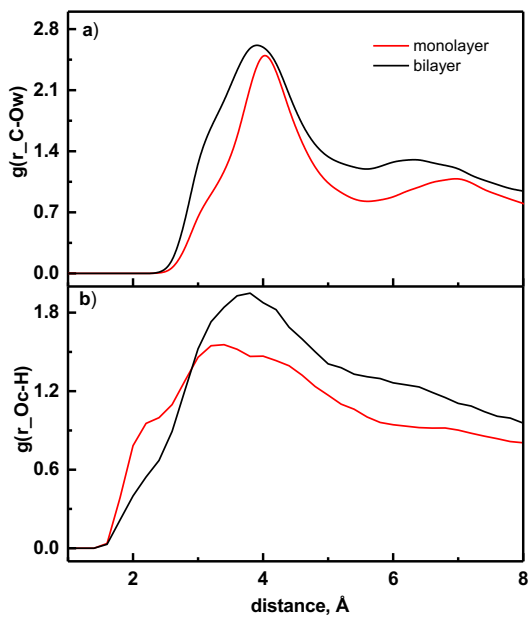

Figure 10: Radial distribution functions for **a)** carbon – water oxygen (water) and **b)** $CO_2$ oxygen – water hydrogen in Na-MMT-$CO_2$-$H_2O$ at $P = 125$ bar and $T = 348.15$ K.

For the monolayer arrangement of Na-MMT-$H_2O$-$CO_2$, the $O_{CO2}$-$H_W$ RDF displays a well-defined shoulder near 2.3 Å, consistent with H-bond formation between $H_2O$ and $CO_2$. This shoulder is much less pronounced for the bilayer. These results are consistent with the calculated average lifetimes[78] of the H bonds between the $O_{CO2}$ atoms and the hydrogen atoms of the water molecules. The values are 1.3 and 0.2 *ps* ($P$ = 125 bar, Na-MMT-$CO_2$-$H_2O$) for the monolayer and the bilayer, respectively, implying stronger H-bonding in the former case. For Na-BEI-$CO_2$-$H_2O$ the corresponding computed lifetimes are 2.2 and 0.4 *ps* for 1W and 2W, respectively. The different lifetimes in the two clays can be related to details of the $CO_2$ distributions in the interlayers.

Figure 11 depicts number density maps for carbon dioxide molecules residing in a selected monolayer or bilayer of Na-MMT-$CO_2$-$H_2O$, in a monolayer of Na-BEI-$CO_2$-$H_2O$, and in pyrophyllite at a $CO_2$/$H_2O$ composition corresponding to that of the monolayer of Na-MMT-$CO_2$-$H_2O$. The maps were obtained by scanning the interlayer space to compute density distributions in planes parallel to the clay surfaces with a step size of 0.1 Å. Then, the distributions were projected onto the *xy* plane and averaged for 10 *ns* of simulation time.

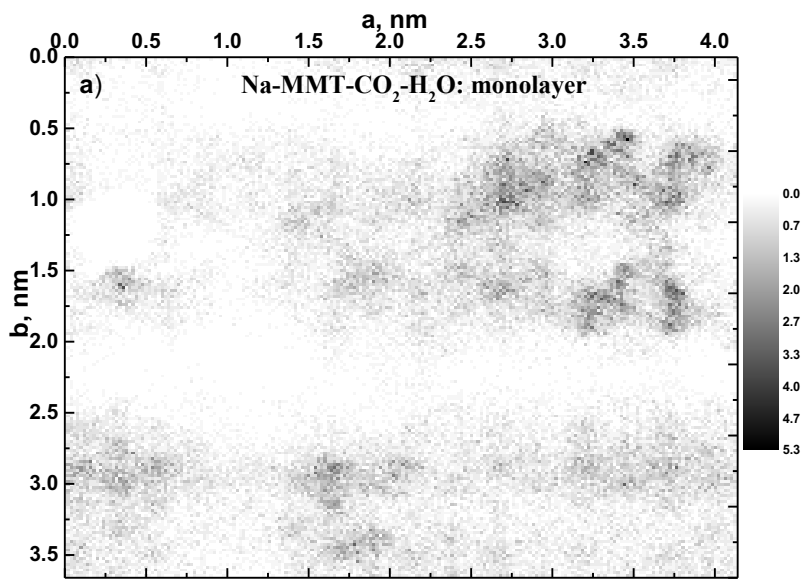

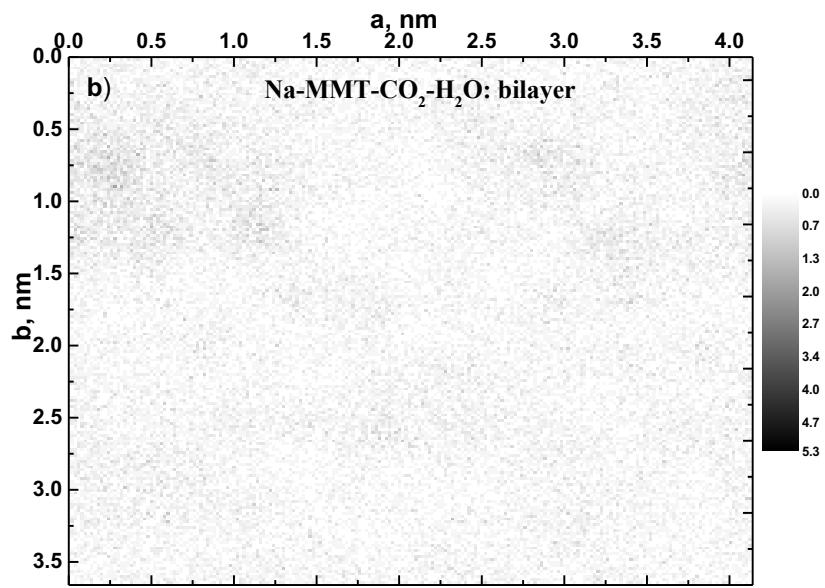

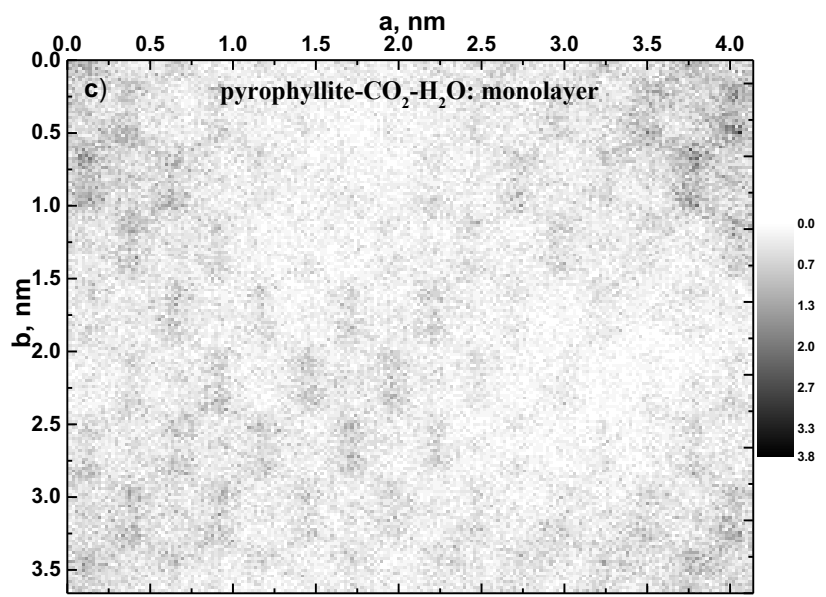

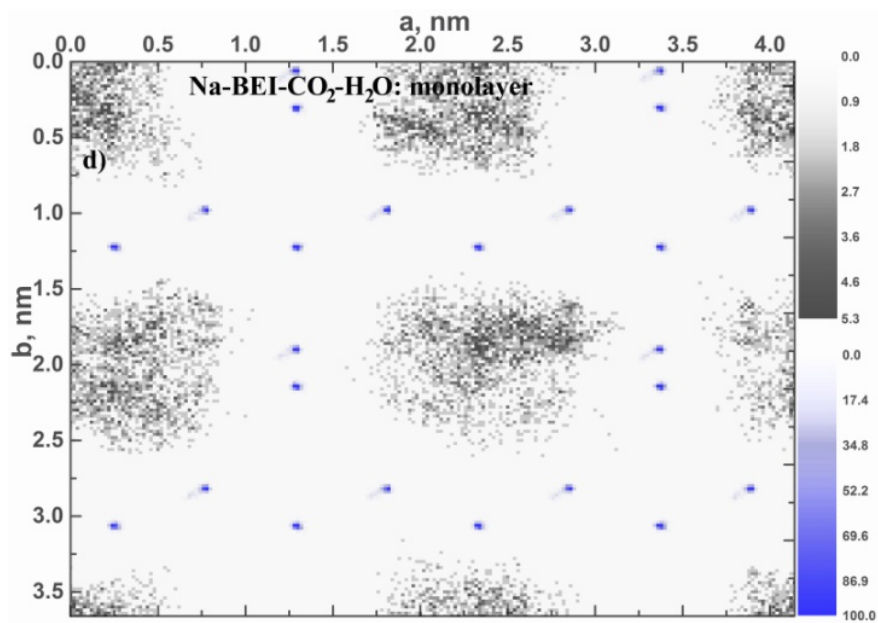

Figure 11. Density maps (number of molecules / nm$^3$) showing carbon dioxide molecule distributions at P = 125 bar and T=348.15 K in the interlayer at $d_{001}$-spacing **a)** 12.5 Å for Na-MMT-CO$_2$-H$_2$O, **b)** 15.5 Å for Na-MMT-CO$_2$-H$_2$O, **c)** 12.5 Å for pyrophyllite, and **d)** 12.5 Å for Na-BEI-CO$_2$-H$_2$O with positions of Al substitutions in the adjacent tetrahedral layers given in blue. Results obtained by averaging over 10 *ns* of simulation time.

Pyrophyllite is a 2:1 dioctahedral phyllosilicate without charges on the mineral layers and, hence, no interlayer cations. The density map for pyrophyllite is included to provide additional insight into the effect of interlayer ions on the CO$_2$ distribution. For Na-MMT-CO$_2$-H$_2$O the distribution of CO$_2$ molecules is distinctly different in the monolayer and bilayer arrangements.

For the monolayer of Na-MMT-CO$_2$-H$_2$O, the density map reveals that the CO$_2$ molecules are non-uniformly distributed with respect to their *XY* coordinates and tend to conglomerate as represented by the dark regions in Figure 11. Analysis of distributions computed over various time frames shows that the clustered carbon dioxide molecules move over time in the interlayer. Migrating CO$_2$ clusters persist even in simulations carried out as long as for 50 *ns*. However, it is expected that averaged over sufficiently long times, the structure would smear out. The formation of CO$_2$ conglomerates in clay monolayers is supported by other studies that speculated that CO$_2$ enters the sub-1W interlayer "by pushing the structural units apart".[4,64] Young and Smith[10] reported that for Sr-MMT under sub 1W conditions the water molecule aggregates around the cations leaving voids. The voids can be filled by carbon dioxide in a "prop-and-fill" mechanism[10] in which the clay is propped open by hydrated cations making possible entrance of $CO_2$. In this work, it is found that the instantaneous spatial distribution of the clustered $CO_2$ molecules in Na-MMT is not correlated with the positions of the substitutions in the $O$ sheets. However, it should be kept in mind that we have used a periodic rather than random distribution of substitution sites in the $T$ or $O$ layers, and that the distribution of $CO_2$ conglomerates could depend on the distribution of substitution sites. Earlier Monte Carlo simulations of cation distributions in the octahedral sheets of phyllosilicates suggest a tendency for Mg cations not to cluster in the Al/Mg systems[79] whereas in some montmorillonite samples $Mg^{2+}$ were observed to form clusters.[80] Palin *et al.*[81] using Monte Carlo simulations show short-range Al-Si ordering across the tetrahedral sheets of muscovite that depends on the Al:Si ratio and temperature. So, the ordering calculated in $T$ sheets of muscovite supports the ordered placement of isomorphic substitutions in $T$ sheets adopted in this work. In a previous study, we also found conglomerate formation of $CO_2$ for both Na- and Ca-MMT.[35] Thus, this behavior is not limited to clays with $Na^+$ ions.

We hypothesize that the interlayer ions facilitate $CO_2$ conglomeration in the monolayer of the clay systems by "tying" up water molecules. To investigate this possibility, a simulation was conducted on pyrophyllite using the equilibrium $CO_2/H_2O$ compositions computed for monolayer of Na-MMT-$CO_2$-$H_2O$ at $P = 125$ bar. The resulting density distribution map, depicted in Figure 11, shows that $CO_2$ develops a periodic pattern reminiscent of ditrigonal rings of the basal surfaces. Thus, in the absence of interlayer ions, the tendency of $CO_2$ to conglomerate is suppressed, and its distribution is determined by the basal atoms of adjacent internal clay surfaces. In a previous paper devoted to turbostratic smectites, we showed that a rotational shift of one adjacent clay layer relative to the other creates Moiré patterns consisting of basal atoms.[35] Those patterns play a role in determining the distribution of interlayer species including carbon dioxide.

The distribution of $CO_2$ in the monolayer of Na-BEI-$CO_2$-$H_2O$ (Figure 11) is particularly striking. The $CO_2$ molecules are clustered in the regions defined by the positions of the isomorphic substitutions in the $T$ sheets (depicted by blue dots in Figure 11). $CO_2$ clearly tends to locate in the areas away from the charge originated from the substitutions, while the interlayer space near substitutions is occupied by water and hydrated cations. In the bilayer, the $CO_2$ density profile displays two-peaks (Figure 7) implying that the $CO_2$ molecules are exposed to the internal surfaces at the comparable distances as in monolayer. Indeed, the corresponding distribution map (not shown) also reveals localized density regions controlled by the substitution positions, however, the structure is more smeared than in the monolayer Obviously, in natural samples the substitution pattern could be irregular, and the $CO_2$ distribution would be more complex than shown in Figure 11. Nonetheless, the location of the layer charges in Na-BEI is a key factor in determining the $CO_2$ distribution in monolayer and bilayer. In Na-BEI-$CO_2$-$H_2O$ the $CO_2$ motion is significantly restricted compared to the more mobile $CO_2$ conglomerates in Na-MMT-$CO_2$-$H_2O$. To quantify the transport properties of the interlayer species, diffusion coefficients were calculated and analyzed for both clays.

### 3.2.2. Diffusion of the interlayer species

Table 2 reports the computed diffusion coefficients for the interlayer species in Na-MMT, Na-BEI, and bulk phases obtained in this work and in other studies[40,82-85] as well as from experimental measurements on similar systems. Both 3D and 2D (xy) values of the diffusion constants are given. In general, the diffusion constants for lateral motion of interlayer species are about 25-35% greater than those computed for 3D motion. As expected, the diffusion coefficients for $CO_2$, $H_2O$, and sodium ions increase in going from monolayer to bilayer arrangements, with the bilayer values being 5-10 times larger than the corresponding monolayer values. For Na-MMT-$CO_2$-$H_2O$ the diffusion coefficients computed in this work are systematically higher than those calculated by Botan et al.[40] (Table 2). This is attributed to differences in the force fields used in the two works. Nevertheless, the calculated values of the diffusion constants of water from the two studies are close to the experimental result obtained using QENS spectroscopy. Also, our calculated diffusion constant for lateral motion of $H_2O$ in Na-MMT is comparable to that measured for water in the interlayer region of Na-hectorite[82,83] (a trioctahedral smectite with isomorphous substitution in the octahedral sheet that displays swelling properties similar to those of MMT).[84]

Table 2: Diffusion coefficients ($\times 10^{-5}$ cm$^2$/s) of water, carbon dioxide molecules, and Na$^+$ cations in the interlayer spaces of Na-MMT and Na-BEI for the equilibrium $CO_2$/$H_2O$ compositions corresponding to $d_{001}$-spacing of 12.5 and 15.5 Å at $T$=347.15 K.

| | | $H_2O$ | | $CO_2$ | | Na$^+$ | |
|---|---|---|---|---|---|---|---|
| | P, bar | D | $D_{xy}$[a] | D | $D_{xy}$[a] | D | $D_{xy}$[a] |
| | | | | Monolayer | | | |
| BEI-$CO_2$-$H_2O$ | 125 | 0.23(1) | 0.32(2) | 0.03(0) | 0.04(0) | 0.01(0) | 0.01(0) |
| | 25 | 0.33(3) | 0.49(3) | 0.05(1) | 0.08(1) | 0.01(0) | 0.02(0) |
| MMT-$CO_2$-$H_2O$ | 125 | 0.41(3) | 0.62(4) | 0.28(3) | 0.43(1) | 0.11(1) | 0.16(2) |
| | 25 | 0.61(2) | 0.91(4) | 0.24(3) | 0.35(4) | 0.20(1) | 0.27(1) |

| System | P | $D_{H_2O}$ | $D_{xy,H_2O}$ | $D_{CO_2}$ | $D_{xy,CO_2}$ | $D_{Na^+}$ | $D_{xy,Na^+}$ |
|---|---|---|---|---|---|---|---|
| Bulk water | 125 | 8.4(7) | | 10.3(8) | | 2.7(5) | |
| MMT-$CO_2$-$H_2O$[40] | 125 | | 0.32(2) | | 0.17(2) | | 0.13(2) |
| | 25 | | 0.51(2) | | 0.24(4) | | 0.19(2) |
| Hectorite-$H_2O$[82] (Expt., T=347K) | 1 | | 1.0(1) | | | | 0.13(2)[85] |
| Bilayer | | | | | | | |
| BEI-$CO_2$-$H_2O$ | 125 | 0.98(4) | 1.48(4) | 0.93(5) | 1.50(6) | 0.10(2) | 0.14(2) |
| | 25 | 1.39(5) | 2.04(7) | 1.29(4) | 2.05(7) | 0.11(3) | 0.17(4) |
| MMT-$CO_2$-$H_2O$ | 125 | 2.39(3) | 3.58(6) | 1.55(6) | 2.50(1) | 1.12(2) | 1.73(2) |
| | 25 | 2.42(4) | 3.63(7) | 1.64(8) | 2.81(5) | 1.38(2) | 1.92(3) |
| MMT-$CO_2$-$H_2O$[40] | 125 | | 3.2(1) | | 1.5(1) | | 1.4(1) |
| | 25 | | 3.5(1) | | 1.6(2) | | 1.3(1) |
| Hectorite-$H_2O$[83] (Expt., T=347K) | 1 | | 2.2(2) | | | | |

$^a D_{xy}$ denotes lateral a diffusion coefficient for motion parallel to clay surfaces.

The calculated diffusion coefficient of water in the bulk phase at is 8.4 x $10^{-5}$ cm$^2$/s, in reasonable agreement with the experimental value of 5.97(6) x $10^{-5}$ cm$^2$/s[86] at T = 347 K. More importantly, the calculated diffusion constant for water in the bulk is 9-28 times greater than that for water in the monolayer and about 2-6 times larger than that of the bilayer for the two clays.[40] For both the monolayer and bilayer cases, the diffusion constant for water in Na-BEI-$CO_2$-$H_2O$ is about two times smaller than for water in Na-MMT-$CO_2$-$H_2O$. Similarly, the measured diffusion constant for $CO_2$ in bulk water, 5.40 ± 0.10 x $10^{-5}$ cm$^2$/s ($T$ = 348 K),[87] is about 1.5-3.0 times larger than for lateral motion of $CO_2$ in the clay bilayer and about an order of magnitude larger than that for lateral diffusion of $CO_2$ in the monolayer of Na-MMT-$CO_2$-$H_2O$. Due to the constrained motion of $CO_2$ in the monolayer of Na-BEI-$CO_2$-$H_2O$, the diffusion coefficient of $CO_2$ in that system is much smaller (by 5-10 times) than for Na-MMT-$CO_2$-$H_2O$.

Our calculated value of the $Na^+$ diffusion constant in bulk water, 2.7 x $10^{-7}$ cm$^2$/s, in reasonable agreement with the experimental value of 3.5 x $10^{-5}$ cm$^2$/s.[88] It is also about twice as large as that for diffusion in $Na^+$ in the bilayer and more than an order of magnitude larger than that for diffusion of $Na^+$ in the monolayer of Na-MMT-$CO_2$-$H_2O$. Predictably, Na-BEI-$CO_2$-$H_2O$, for which the $Na^+$ ions tend to be adsorbed at the internal surfaces, has much smaller $Na^+$ diffusion constants than for Na-MMT-$CO_2$-$H_2O$ in both the monolayer and bilayer cases. Kozaki et al.[89] reported a diffusion constant of 1.8x$10^{-6}$ cm$^2$/s for interlayer sodium ions in the 2W hydration state of MMT at 323 K and 0.1-0.3 M concentrations of NaCl. Even though that value is in good agreement with that calculated in the present study for the 2W state of Na-MMT, direct comparison between calculated and measured ionic mobilities of smectite samples should be done with caution since the experimental values can include substantial contributions from ions residing in pores and at external surfaces depending on experimental conditions and ionic strength.[89]

## IV. Conclusions

Molecular dynamics and multiphase Gibbs ensemble Monte Carlo simulations were carried out to study adsorption of carbon dioxide and water in the interlayer regions of Na-MMT and Na-BEI. The free energy calculations indicate that intercalation of pure $CO_2$ into the clay interlayers is not favorable, while pure $H_2O$ adsorption may occur naturally at the ($P$, $T$) conditions relevant to carbon sequestration in deep geological formations. For Na-MMT-$CO_2$-$H_2O$ the free energy of swelling shows two minima corresponding to monolayer and bilayer arrangements of the interlayer species, whereas for Na-BEI-$CO_2$-$H_2O$ the expansion of the interlayer is limited to monolayer for both pure water and $CO_2$/water mixtures. This is a consequence of the strong interaction between the cations and the negatively charged clay layers.

The calculations predict the carbon dioxide concentrations in the interlayer of both clays greatly exceeding the solubility of $CO_2$ in bulk water at the ($P$, $T$) conditions of interest. In agreement with experiment, the maximum carbon dioxide adsorption occurs at the $d_{001}$-spacing range (11.5-12.5 Å) corresponding to the 1W hydration state. For both clays, $CO_2$ can fill the available space in sub-1W interlayer arrangements, resulting in expansion of $d_{001}$ to 12.5 Å. The presence of interlayer cations leads to $CO_2$ conglomeration in the monolayer of Na-MMT-$CO_2$-$H_2O$. For Na-BEI-$CO_2$-$H_2O$, both the cations and the negative charges in the tetrahedral layers significantly impact the distributions of water and carbon dioxide molecules. $CO_2$ is driven away from the positions of isomorphic substitutions, whereas hydrated cations molecules are located close to $Al^{3+}$/$Si^{4+}$ substitution sites. The results showed that in smectites, hydrated cations and substitutions into tetrahedral sheets promote segregation of water and $CO_2$ in the interlayers and affects the ability of the clay minerals to retain carbon dioxide.

Not surprisingly, in the monolayer region of smectite clays, the mobilities of all interlayer species: water, $CO_2$, and $Na^+$ are significantly retarded compared to the bulk phase. The corresponding diffusion constants in the bilayer regions are significantly less reduced compared to their values in bulk water. The most prominent reduction in mobility is found for $CO_2$ and $Na^+$ in Na-BEI-$CO_2$-$H_2O$.

The simulations predict that at equilibrium with a $CO_2$-bearing aquifer, carbon dioxide intercalation into hydrated clay phases is accompanied by expansion of the interlayer distance causing increase of clay volume, which, in turn, influences the porosity and permeability of formations enriched in expandable clay minerals. The greater solubility of $CO_2$ in the interlayer region of smectites than in bulk water suggests that smectite minerals are good candidates for carbon dioxide storage.

**Acknowledgment**

This technical effort was performed in support of the National Energy Technology Laboratory's ongoing research in Subtask 4000.4.641.061.002.254 under the RES contract DE-FE0004000. The simulations were carried out on NETL's High-Performance Computer for Energy and the Environment (HPCEE) and on computers in the University of Pittsburgh's Center for Simulation and Modeling.

# References

(1) Houghton, J. Global Warming: The Complete Briefing; Cambridge University Press: Cambridge, U.K., 2009.

(2) Benson, S. M.; Cole, D. R. $CO_2$ sequestration in deep sedimentary formations. *Elements* **2008**, *4*, 325-331.

(3) Giesting, P.; Guggenheim, S.; Koster van Groos, A. F.; Busch, A. Interaction of carbon dioxide with Na-exchanged montmorillonite at pressures to 640 bars: implications for $CO_2$ sequestration. *Int. J. Greenhouse Gas Control* **2012**, *8*, 73-81.

(4) Giesting, P.; Guggenheim, S.; Koster van Groos, A. F.; Busch, A. X-ray diffraction study of K- and Ca-exchanged montmorillonites in $CO_2$ atmospheres. *Environ. Sci. Technol.* **2012**, *46*, 5623-5630.

(5) Rother, G.; Ilton, E. S.; Wallacher, D.; Hauβ, T.; Schaef, H. T.; Qafoku, O.; Rosso, K. M.; Felmy, A. R.; Krukowski, E. G.; Stack, A. G.; et al. $CO_2$ sorption to subsingle hydration layer montmorillonite clay studied by excess sorption and neutron diffraction measurements. *Environ. Sci. Technol.* **2012**, *47*, 205-211.

(6) Ilton, E. S.; Schaef, H. T.; Qafoku, O.; Rosso, K. M.; Felmy, A. R. In situ X-ray diffraction study of $Na^+$ saturated montmorillonite exposed to variably wet super critical $CO_2$. *Environ. Sci. Technol.* **2012**, *46*, 4241-4248.

(7) Norrish, K. The swelling of montmorillonite. *Discuss. Faraday Soc.* **1954**, *18*, 120-134.

(8) Hensen, E. J.; Smit, Why clays swell. B. *J. Phys. Chem. B* **2002**, *106*, 12664-12667.

(9) Berend, I.; Cases, J.; Francois, M.; Uriot, J.; Michot, L.; Masion, A.; Thomas, F. Mechanism of adsorption and desorption of water vapor by homoionic montmorillonites: 2. the Li, Na, K, Rb and $Cs^+$-exchanged forms. *Clays Clay Miner.* **1995**, *43*, 324-336.

(10) Young, D. A.; Smith, D. E. Simulations of clay mineral swelling and hydration: dependence upon interlayer ion size and charge. *J. Phys. Chem. B* **2000**, *104*, 9163-9170.

(11) Douillard, J.-M.; Lantenois, S.; Prelot, B.; Zajac, J.; Henry, M. Study of the influence of location of substitutions on the surface energy of dioctahedral smectites. *J. Colloid Interface Sci.* **2008**, *325*, 275-281.

(12) Skipper, N.; Refson, K.; McConnell, J. Computer simulation of interlayer water in 2:1 clays. *J. Chem. Phys.* **1991**, *94*, 7434-7445.

(13) Skipper, N.; Chang, F.-R. C.; Sposito, G. Monte Carlo simulation of interlayer molecular structure in swelling clay minerals. i: Methodology. *Clays Clay Miner.* **1995**, *43*, 285-293.

(14) Skipper, N.; Sposito, G.; Chang, F.-R. C. Monte Carlo simulation of interlayer molecular structure in swelling clay minerals. 2. Monolayer hydrates. *Clays Clay Miner.* **1995**, *43*, 294-303.

(15) Delville, A. Modeling the clay-water interface. *Langmuir* **1991**, *7*, 547-555.

(16) Delville, A. Structure of liquids at a solid interface: an application to the swelling of clay by water. *Langmuir* **1992**, *8*, 1796-1805.

(17) Delville, A. Structure and properties of confined liquids: a molecular model of the clay-water interface. *J. Phys. Chem.* **1993**, *97*, 9703-9712.

(18) Chang, F.-R. C.; Skipper, N.; Sposito, G. Computer simulation of interlayer molecular structure in sodium montmorillonite hydrates. *Langmuir* **1995**, *11*, 2734-2741.

(19) Bordarier, P.; Rousseau, B.; Fuchs, A. H. Rheology of model confined ultrathin fluid films. i. statistical mechanics of the surface force apparatus experiments. *J. Chem. Phys.* **1997**, *106*, 7295-7302.

(20) Marry, V.; Turq, P.; Cartailler, T.; Levesque, D. Microscopic simulation of structure and dynamics of water and counterions in a monohydrated montmorillonite. *J. Chem. Phys.* **2002**, *117*, 3454-3463.

(21) Suter, J. L.; Sprik, M.; Boek, E. S. Free energies of absorption of alkali ions onto beidellite and montmorillonite surfaces from constrained molecular dynamics simulations. *Geochim. Cosmochim. Acta* **2012**, *91*, 109-119.

(22) Teppen, B. J.; Rasmussen, K.; Bertsch, P. M.; Miller, D. M.; Schaefer, L. Molecular dynamics modeling of clay minerals. 1. Gibbsite, kaolinite, pyrophyllite, and beidellite. *J. Phys. Chem. B* **1997**, *101*, 1579-1587.

(23) Greathouse, J. A.; Cygan, R. T. Water structure and aqueous uranyl (VI) adsorption equilibria onto external surfaces of beidellite, montmorillonite, and pyrophyllite: Results from molecular simulations. *Environ. Sci. Technol.* **2006**, *40*, 3865-3871.

(24) Fu, M.; Zhang, Z.; Low, P. Changes in the properties of a montmorillonite-water system during the adsorption and desorption of water: hysteresis. *Clays Clay Miner.* **1990**, *38*, 485-492.

(25) Cygan, R. T.; Liang, J.-J.; Kalinichev, A. G. Molecular models of hydroxide, oxyhydroxide, and clay phases and the development of a general force field. *J. Phys. Chem. B* **2004**, *108*, 1255-1266.

(26) Smith, D. E. Molecular computer simulations of the swelling properties and interlayer structure of cesium montmorillonite. *Langmuir* **1998**, *14*, 5959-5967.

(27) Mooney, R.; Keenan, A.; Wood, L. Adsorption of water vapor by montmorillonite. i. Heat of desorption and application of BET theory. *J. Am. Chem. Soc.* **1952**, *74*, 1367-1371.

(28) Calvet, R. Hydratation de la Montmorillonite et Diffusion des Cations Compensateurs. Université VI, Faculté des Sciences Physiques, 1972.

(29) Whitley, H. D.; Smith, D. E. Free energy, energy, and entropy of swelling in Cs-, Na-, and Sr-montmorillonite clays. *J. Chem. Phys.* **2004**, *120*, 5387-5395.

(30) Schaef, H. T.; Ilton, E. S.; Qafoku, O.; Martin, P. F.; Felmy, A. R.; Rosso, K. M. In situ XRD study of $Ca^{2+}$-saturated montmorillonite (Stx-1) exposed to anhydrous and wet supercritical carbon dioxide. *Int. J. Greenhouse Gas Control* **2012**, *6*, 220-229.

(31) Odriozola, G.; Aguilar, J.; Lopez-Lemus, J. Na-montmorillonite hydrates under ethane rich reservoirs: $NP_{zz}T$ and $\mu P_{zz}T$ simulations. *J. Chem. Phys.* **2004**, *121*, 4266-4275.

(32) Billemont, P.; Coasne, B.; De Weireld, G. An experimental and molecular simulation study of the adsorption of carbon dioxide and methane in nanoporous carbons in the presence of water. *Langmuir* **2010**, *27*, 1015-1024.

(33) Ferrage, E.; Sakharov, B. A.; Michot, L. J.; Delville, A.; Bauer, A.; Lanson, B.; Grangeon, S.; Frapper, G.; Jim´enez-Ruiz, M.; Cuello, G. J. Hydration properties and interlayer organization of water and ions in synthetic Na-smectite with tetrahedral layer charge. Part 2. Toward a precise coupling between molecular simulations and diffraction data. *J. Phys. Chem. C* **2011**, *115*, 1867-1881.

(34) Zhang, G.; Al-Saidi, W. A.; Myshakin, E. M.; Jordan, K. D. Dispersion-corrected density functional theory and classical force field calculations of water loading on a pyrophyllite(001) surface. *J. Phys. Chem. C* **2012**, *116*, 17134-17141.

(35) Cygan, R. T.; Romanov, V. N.; Myshakin, E. M. Molecular simulation of carbon dioxide capture by montmorillonite using an accurate and flexible force field. *J. Phys. Chem. C* **2012**, *116*, 13079-13091.

(36) Myshakin, E. M.; Saidi, W. A.; Romanov, V. N.; Cygan, R. T.; Jordan, K. D. Molecular dynamics simulations of carbon dioxide intercalation in hydrated Na-montmorillonite. *J. Phys. Chem. C* **2013**, *117*, 11028-11039.

(37) Churakov, S. V. Mobility of Na and Cs on montmorillonite surface under partially saturated conditions. *Environ. Sci. Technol.* **2013**, *47*, 9816-9823.

(38) Myshakin, E.; Makaremi, M.; Romanov, V. N.; Jordan, K. D.; Guthrie, G. D. Molecular dynamics simulations of turbostratic dry and hydrated montmorillonite with intercalated carbon dioxide. *J. Phys. Chem. A* **2014**, *118*, 7454-7468.

(39) Titiloye, J.; Skipper, N. Computer simulation of the structure and dynamics of methane in hydrated Na-smectite clay. *Chem. Phys. Lett.* **2000**, *329*, 23-28.

(40) Titiloye, J. O.; Skipper, N. T. Monte Carlo and molecular dynamics simulations of methane in potassium montmorillonite clay hydrates at elevated pressures and temperatures. *J. Colloid Interface Sci.* **2005**, *282*, 422-427.

(41) Liu, X.-D.; Lu, X.-C. A thermodynamic understanding of clay-swelling inhibition by potassium ions. *Angewandte Chemie International Edition* **2006**, *45*, 6300-6303.

(42) Rotenberg, B.; Morel, J.-P.; Marry, V.; Turq, P.; Morel-Desrosiers, N. On the driving force of cation exchange in clays: Insights from combined microcalorimetry experiments and molecular simulation. *Geochim. Cosmochim. Acta* **2009**, *73*, 4034-4044.

(43) Botan, A.; Rotenberg, B.; Marry, V.; Turq, P.; Noetinger, B. Carbon dioxide in montmorillonite clay hydrates: thermodynamics, structure, and transport from molecular simulation. *J. Phys. Chem. C* **2010**, *114*, 14962-14969.

(44) Panagiotopoulos, A. Z. Adsorption and capillary condensation of fluids in cylindrical pores by Monte Carlo simulation in the Gibbs ensemble. *Mol. Phys.* **1987**, *62*, 701-719.

(45) McGrother, S. C.; Gubbins, K. E. Constant pressure Gibbs ensemble Monte Carlo simulations of adsorption into narrow pores. *Mol. Phys.* **1999**, *97*, 955-965.

(46) Diestler, D.; Schoen, M.; Curry, J. E.; Cushman, J. H. Thermodynamics of a fluid confined to a slit pore with structured walls. *J. Chem. Phys.* **1994**, *100*, 9140-9146.

(47) Lopes, J. C.; Tildesley, D. Multiphase equilibria using the Gibbs ensemble Monte Carlo method. *Mol. Phys.* **1997**, *92*, 187-196.

(48) Martin, M. G.; Siepmann, J. I. Calculating Gibbs free energies of transfer from Gibbs ensemble Monte Carlo simulations. *Theor. Chem. Acc.* **1998**, *99*, 347-350.

(49) Liu, A.; Beck, T. L. Vapor-liquid equilibria of binary and ternary mixtures containing methane, ethane, and carbon dioxide from Gibbs ensemble simulations. *J. Phys. Chem. B* **1998**, *102*, 7627-7631.

(50) Kristof, T.; Vorholz, J.; Maurer, G. Molecular simulation of the high-pressure phase equilibrium of the system carbon dioxide-methanol-water. *J. Phys. Chem. B* **2002**, *106*, 7547-7553.

(51) Carrero-Mantilla, J.; Llano-Restrepo, M. Vapor-liquid equilibria of the binary mixtures nitrogen+methane, nitrogen+ethane and nitrogen+carbon dioxide, and the ternary mixture nitrogen+methane+ethane from Gibbs-ensemble molecular simulation. *Fluid Phase Equilib.* **2003**, *208*, 155-169.

(52) Zhang, L.; Siepmann, J. I. Pressure dependence of the vapor-liquid-liquid phase behavior in ternary mixtures consisting of n-alkanes, n-perfluoroalkanes, and carbon dioxide. *J. Phys. Chem. B* **2005**, *109*, 2911-2919.

(53) Loewenstein, W. The distribution of aluminum in the tetrahedra of silicates and aluminates. *Am. Mineral.* **1954**, *39*, 92-96.

(54) Berendsen, H. J.; Postma, J. P.; Van Gunsteren, W. F.; Hermans, J. Interaction models for water in relation to protein hydration. Intermolecular Forces: Springer 1981, 331-342.

(55) Allen, M. P.; Tildesley, D. J. Computer Simulation of Liquids: Clarendon, Oxford, 1989.

(56) Wiebe, R. The binary system carbon dioxide-water under pressure. *Chem. Rev.* **1941**, *29*, 475-481.

(57) Gillespie, P. C.; Wilson, G. M. Vapor-Liquid and Liquid-Liquid Equilibria: Water- Methane, Water-Carbon Dioxide, Water-Hydrogen Sulfide, Water-Npentane, Water-Methane-Npentane: Gas Processors Association, 1982.

(58) Vorholz, J.; Harismiadis, V.; Rumpf, B.; Panagiotopoulos, A.; Maurer, G. Vapor+liquid equilibrium of water, carbon dioxide, and the binary system, water+carbon dioxide, from molecular simulation. *Fluid Phase Equilib.* **2000**, *170*, 203-234.

(59) McDonald, I. NPT-ensemble Monte Carlo calculations for binary liquid mixtures. *Mol. Phys.* **1972**, *23*, 41-58.

(60) Mooij, G.; Frenkel, D.; Smit, B. Direct simulation of phase equilibria of chain molecules. *J. Phys.: Condens. Matter* **1992**, *4*, L255.

(61) Martin, M. G.; Frischknecht, A. L. Using arbitrary trial distributions to improve intramolecular sampling in configurational-bias Monte Carlo. *Mol. Phys.* **2006**, *104*, 2439-2456.

(62) Frenkel, D.; Smit, B. Understanding Molecular Simulation: From Algorithms to Applications: Academic press, 2001; Vol. 1.

(63) MCCCS Towhee version 7.0.6. http://towhee.sourceforge.net, Accessed: 2014-07-1.

(64) Nóse, S. A molecular dynamics method for simulations in the canonical ensemble. *Mol. Phys.* **1984**, *52*, 255-268.

(65) Hoover, W. G. Canonical dynamics: equilibrium phase-space distributions. *Phys. Rev. A* **1985**, *31*, 1695.

(66) Van Gunsteren, W.; Berendsen, H. A leap-frog algorithm for stochastic dynamics. *Mol. Simul.* **1988**, *1*, 173-185.

(67) Van Der Spoel, D.; Hess B.; Lindahl E. The GROMACS development team, GROMACS

User Manual version 4.6.5, www.gromacs.org. 2013. Accessed: 2014-09-28.

(68) Loring, J. S.; Schaef, H. T.; Turcu, R. V.; Thompson, C. J.; Miller, Q. R.; Martin, P. F.; Hu, J.; Hoyt, D. W.; Qafoku, O.; Ilton, E. S.; et al. In situ molecular spectroscopic evidence for $CO_2$ intercalation into montmorillonite in supercritical carbon dioxide. *Langmuir* **2012**, *28*, 7125-7128.

(69) Cases, J.; Berend, I.; Besson, G.; Francois, M.; Uriot, J.; Thomas, F.; Poirier, J. Mechanism of adsorption and desorption of water vapor by homoionic montmorillonite. 1. The sodium-exchanged form. *Langmuir* **1992**, *8*, 2730-2739.

(70) Karaborni, S.; Smit, B.; Heidug, W.; Urai, J.; Van Oort, E. The swelling of clays: molecular simulations of the hydration of montmorillonite. *Science* **1996**, *271*, 1102.

(71) Chavez-Paez, M.; Van Workum, K.; De Pablo, L.; De Pablo, J. J. Monte Carlo simulations of Wyoming sodium montmorillonite hydrates. *J. Chem. Phys.* **2001**, *114*, 1405-1413.

(72) Anderson, R.; Ratcli e, I.; Greenwell, H.; Williams, P.; Cli e, S.; Coveney, P. Clay swelling: a challenge in the oil field. *Earth-Sci. Rev.* **2010**, *98*, 201-216.

(73) Smith, D. E.; Wang, Y.; Whitley, H. D. Molecular simulations of hydration and swelling in clay minerals. *Fluid Phase Equilib.* **2004**, *222*, 189-194.

(74) Kloprogge, J. T.; Jansen, J. B. H.; Geus, J. W. Characterization of synthetic Na-beidellite. *Clays Clay Miner.* **1990**, *38*, 409-414.

(75) Loring, J. S.; Ilton, E. S.; Chen, J.; Thompson, C. J.; Martin, P. F.; Benezeth, P.; Rosso, K. M.; Felmy, A. R.; Schaef, H. T. In situ study of $CO_2$ and $H_2O$ partitioning between Na-montmorillonite and variably wet supercritical carbon dioxide. *Langmuir* **2014**, *30*, 6120-6128.

(76) Ansell, S.; Barnes, A.; Mason, P.; Neilson, G.; Ramos, S. X-ray and neutron scattering studies of the hydration structure of alkali ions in concentrated aqueous solutions. *Biophys. Chem.* **2006**, *124*, 171-179.

(77) Soper, A. K.; Weckstrom, K. Ion solvation and water structure in potassium halide aqueous solutions. *Biophys. Chem.* **2006**, *124*, 180-191.

(78) Khalack, J.; Lyubartsev, A. Car-Parrinello molecular dynamics simulations of $Na^+$- $Cl^-$ ion pair in liquid water. *Condens. Matter Phys.* **2004**, *7*, 683-698.

(79) Palin, E.J.; Dove, M. T.; Hernández-Laguna, A.; Santz-Díaz, A. A computational investigation of the Al/Fe/Mg order-disorder behavior in the dioctahedral sheet of phyllosilicates. *Am. Mineral.*, **2004**, *89*, 164-175.

(80) Muller, F.; Besson, G.; Manceau, A.; Drits, V. A. Distribution of isomorphous cations within

octahedral sheets in montmorillonite from Camp-Bertaux. *Phys. Chem. Miner*, **1997**, *24*, 159-166.

(81) Palin, E.J.; Dove, M. T.; Redfern, S. A.T.; Bosenick, A.; Santz-Díaz, C. I.; Warren, M. C. Computational study of tetrahedral Al-Si ordering in muscovite. *Phys. Chem. Miner,* **2001**, *28*, 534-544.

(82) Marry, V.; Dubois, E.; Malikova, N.; Durand-Vidal, S.; Longeville, S.; Breu, J. Water dynamics in hectorite clays: influence of temperature studied by coupling neutron spin echo and molecular dynamics. *Environ. Sci. Technol.* **2011**, *45*, 2850-2855.

(83) Marry, V.; Dubois, E.; Malikova, N.; Breu, J.; Haussler, W. Anisotropy of water dynamics in clays: insights from molecular simulations for experimental QENS analysis. *J. Phys. Chem. C* **2013**, *117*, 15106-15115.

(84) Malikova, N.; Cadene, A.; Dubois, E.; Marry, V.; Durand-Vidal, S.; Turq, P.; Breu, J.; Longeville, S.; Zanotti, J. M. Water diffusion in a synthetic hectorite clay studied by quasi-elastic neutron scattering. *J. Phys. Chem. C* **2007**, *111*, 17603-17611.

(85) Marry, V.; Malikova, N.; Cadene, A.; Dubois, E.; Durand-Vidal, S.; Turq, P.; Breu, J.; Longeville, S.; Zanotti, J. M. Water diffusion in a synthetic hectorite by neutron scattering-beyond the isotropic translational model. *J. Phys.: Condens. Matter* **2008**, *20*, 104205.

(86) Holz, M.; Heil, S. R.; Sacco, A. Temperature-dependent self-diffusion coefficients of water and six selected molecular liquids for calibration in accurate $^1$H NMR PFG measurements. *Phys. Chem. Chem. Phys.* **2000**, *2*, 4740-4742.

(87) Lindeberg, E.; Wessel-Berg, D. Vertical convection in an aquifer column under a gas cap of $CO_2$. *Energy Convers. Manage.* **1997**, *38*, S229- S234.

(88) Stokes, R.; Robinson, R. Electrolyte solutions. *Butterworths, London* **1959**.

(89) Kozaki, T.; Fujishima, A.; Sato, S.; Ohashi, H. Self-diffusion of sodium ions in compacted sodium montmorillonite. *Nucl. Technol.* **1998**, *121*, 63-69.

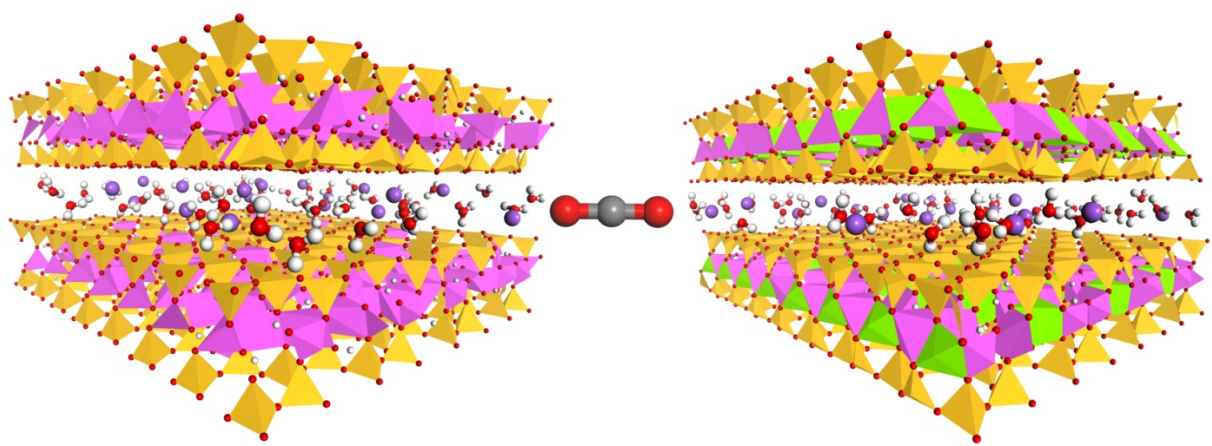

TOC figure. Beidellite (left) and Montmorillonite (right) structures with intercalated water molecules and sodium ions. Carbon dioxide molecule shown in the middle. The atom color coding: purple, red, white, grey, yellow, pink and green for Na, O, H, C, Si, Al and Mg, respectively.